\documentstyle[12pt,aaspp4]{article}

\begin{document}
%\draft
\title{Gamma Ray Bursts from Baryon Decay in Neutron Stars}
\author{Ue-Li Pen and Abraham Loeb}
\affil{Harvard-Smithsonian Center
for Astrophysics, 60 Garden St., Cambridge, MA 02138, USA}
\author{and Neil Turok}
\affil{DAMTP, Cambridge University, Cambridge, United Kingdom}

\newcommand{\etal}{{et al. }}

\begin{abstract}

The standard unbroken electroweak theory is known to erase baryon number.
The baryon number symmetry can be restored in the core of a neutron star as
its density diverges via gravitational instability due to a binary merger
event.  We argue that for certain double Higgs models with discrete
symmetries, this process may result in an expanding self-sustained burning
front which would convert the entire neutron matter into radiation.  This
process would release $\sim 10^{54}~{\rm ergs}$ in electromagnetic
radiation over $\sim 10^{-4}~{\rm sec}$, with negligible baryonic
contamination. The resulting fireball would have all the properties
necessary to produce a $\gamma$-ray burst as a result of its interaction
with ambient interstellar gas.  The subsequent Higgs decay would produce a
millisecond burst of $\sim 10^{52}$ ergs in $\sim 100~{\rm GeV}$ neutrinos
which might be observable.  The above mechanism may have also caused
electroweak baryogenesis in the early universe, giving rise to the observed
matter-antimatter asymmetry today.

\end{abstract}

\keywords{cosmology: theory -- gamma rays: bursts}

%\centerline{submitted to {\it ApJ Letters}, December 1997}

\newcommand{\be}{\begin{equation}}
\newcommand{\ee}{\end{equation}}
\newcommand{\grb}{$\gamma$-ray burst}

\section{Introduction}

The energy source for $\gamma$-ray bursts (GRB) has been enigmatic since
their discovery (see Fishman \& Meegan 1995, for a recent review).  The
recent detection of absorption lines at a redshift $z=0.835$ in the optical
afterglow spectrum of the event GRB970508 (Metzger et al.  1997),
established a firm lower bound on its distance, and hence on its total
radiation energy $\ga 10^{52}$ erg (Waxman 1997).  The long duration of
some GRBs is difficult to reconcile with their millisecond time structure.
If the variability is caused by an external shock due to the interaction of
the fireball with an ambient medium (M\'esz\'aros
\& Rees 1993a,b), then the conversion
efficiency must be small (Sari \& Piran 1997a) and the total energy
released approaches the rest mass energy of a neutron star.  A further
challenge to GRB models is the baryon contamination problem.  The high
Lorentz factors implied by the short variability and non-thermal spectra of
GRBs (see, e.g. Woods \& Loeb 1995) require that the baryonic fraction be
$\la 1\%$, of the total energy.  This constraint is particularly difficult
to satisfy in stellar environments such as neutron star mergers (Ruffert et
al. 1997, and references therein), where a considerable amount of baryonic
debris is unavoidable.

The standard electroweak theory allows for baryon number violating
processes ('t Hooft 1976).  Through a combination of chirality and
topology, baryon and lepton number may be violated.  A change in the
winding number of the gauge fields results in fermions being pulled out or
pushed into the Dirac sea.  Normally, in the broken symmetry vacuum, such
effects are very strongly suppressed since the gauge fields are massive and
difficult to excite. However, in the unbroken phase it is well established
that when the electroweak symmetry is restored, the baryon and lepton number
violation processes are unsuppressed (for a review, see Rubakov \&
Shaposhnikov 1996).  It is natural to ask whether these intriguing
processes could be excited in the Universe today. There are two fundamental
barriers which make their appearance rare: (i) a large energy input is
required to change the gauge/Higgs winding numbers, and (ii) the energy
must be input in a coherent fashion into the long wavelength components of
the bosonic fields (and this, for example, does not happen in high energy
collisions of single elementary particles).

In this {\it Letter} we argue that baryon number symmetry might be
restored in the core of a neutron star if its density is raised to
extreme values.  Such a process might be triggered during the
catastrophic gravitational collapse caused by some cataclysmic event
(such as binary coalescence or substantial accretion), which raises
the neutron star mass above its stability limit.  Once the symmetry is
restored in a sufficiently large (macroscopic) volume, a
self-sustained burning front would propagate throughout the star and
convert the entire neutron matter into radiation, releasing $\sim
1~{\rm GeV}$ per baryon in the process. The burning front will be
accompanied by an electroweak domain wall in which the barrier to
baryon number violation is greatly reduced.  The existence of dense
nuclear matter allows long wavelength modes of the gauge fields to be
naturally excited. Overall, the burning front could release $\sim
10^{54}~{\rm ergs}$ in electromagnetic radiation with negligible
baryonic contamination. The resulting fireball would have all the
properties necessary to produce a GRB due to its interaction with
ambient interstellar gas.

In \S \ref{sec:selfsim} we present the hydrodynamic solution for the
associated electroweak burning front.  We then discuss the trigger
mechanism for the explosive process described above, in \S \ref{sec:ew}.
\S \ref{sec:energ} considers the energetics of electroweak baryon number
violation on moving walls.  Finally, 
we examine the implications of our model for cosmological 
baryogenesis and electroweak theory in \S
\ref{sec:cosm}, and summarize its
testable predictions for GRBs in \S \ref{sec:conc}.

%We argue in Section \S \ref{sec:energ} that sufficient magnetic and
%electric fields may accrue on the propagating domain walls to allow
%baryon number violation to annihilate most of the neutrons the walls
%sweep up. The burning of neutrons into antineutrinos releases energy
%and heats up the matter behind the wall: the resultant pressure drives
%the bubble wall outwards until it annihilates the whole of the neutron
%star.

\section{Self-Similar Explosion}
\label{sec:selfsim}

In our model, we consider three regions of the dynamical explosion.  In the
outer region, the cold neutron star material is at rest.  The shock front
propagates through this material at some speed $\lambda_s c$, where $c$ is
the speed of light.  This burning front is the second region under
consideration, which has a characteristic thickness of an electroweak
length $\sim 10^{-16}$ cm.  The burning front is sustained by the presence
of an electroweak domain wall.  While the microphysics in this region is
complicated, we can describe its hydrodynamic implications as if it were
infinitely thin.  We will return to its discussion in \S \ref{sec:energ}.
The global conservation of energy and momentum thus
dictate the global dynamics of this burning region.  The last region under
consideration includes the postshock (burnt) radiation phase.  This heated
region will initially expand due to the increase in pressure.  This
pressure is directly exerted against the domain wall, which then propagates
further outward.  The goal of this section is to quantify the
self-sustained dynamics of these regions, assuming that the process has
already been triggered. The possible trigger mechanism will be considered
in the next section.

The equations of motion of a relativistic fluid are given by the
energy-momentum conservation equations $T^{\mu\nu}_{;\mu}=0$ where
$T^{\mu\nu}=(\rho+p)u^\mu u^\nu+pg^{\mu\nu}$.  We use a metric signature
convention of $(-,+,+,+)$.  In the rest frame of the star, we have
\begin{eqnarray}
\frac{\partial}{\partial t}
\frac{\rho+p\beta^2}{1-\beta^2}+\frac{1}{r^2}\frac{\partial}{\partial
r} \frac{r^2(\rho+p)\beta}{1-\beta^2} &=& 0 ,
\label{eqn:cont}
\nonumber\\
\frac{\partial}{\partial t}
\frac{(\rho+p)\beta}{1-\beta^2}+\frac{1}{r^2}\frac{\partial}{\partial
r} \frac{r^2(\rho+p)\beta^2}{1-\beta^2} +\frac{\partial p}{\partial r}
&=& 0,
\label{eqn:momt}
\end{eqnarray}
where $r$ is the radial coordinate and the speed of light is set to unity.
We ignore gravity for now, assuming that the star is initially in
static equilibrium, and that the final phase is gravitationally fully
unbound.  

We assume that the pre-shocked neutron star material is described by an
equation of state $p_0 = \alpha\rho_0$.  The pre-shock density 
of a neutron star, $\rho_0\sim 2\times 10^{14}$ g/cm$^3$, is slightly above
nuclear density.  The pressure is typically small, $\alpha
\sim 0.1$, but can reach relativistic values near the center of the
star, where $\alpha\sim 1/3$.  The burnt phase consists of radiation
and electroweak false vacuum energies, $\rho_{\rm
burnt}=\rho_F+\rho_V$, $p_{\rm burnt}=\rho_F/3-\rho_V$.  Typically
$\rho_V \sim \rho_0/100$, as discussed further in \S
\ref{sec:cosm}. 
The burning front has a width of order the electroweak scale, $\sim
10^{-16}$ cm, much smaller than the macroscopic scale of the neutron
star, and can therefore be approximated as a sharp discontinuity.
Since the neutron star density is nearly constant (Shapiro \&
Teukolsky 1983), the explosion should be well approximated by a
self-similar solution (see examples in Blandford \& McKee 1976; Pen
1994).  The amount of energy released per baryon equals to its rest
mass and therefore the Lorentz factor of the burning front cannot be
much greater than unity. This implies that any self-similar
self-sustained burning front (which does not decay as it expands) must
propagate at a constant speed through the neutron matter.

We therefore seek solutions to (\ref{eqn:cont}) and (\ref{eqn:momt}) which
depend on $r,t$ only through the combination $\lambda=r/t$.  We define the
dimensionless densities ${\tilde
\rho}=\rho/\rho_0$ and ${\tilde \rho}_V=V(\phi^-)/\rho_0$, 
where $V$ is defined in \S \ref{sec:cosm} below.  After some algebra, we find
that the equation for $\beta(\lambda)$ separates from the equation for
${\tilde \rho}$:
\be
\lambda\beta'= \frac{2\beta( 1 - {{\beta}^2} ) 
      ( 1 - \beta\lambda ) 
        }{  \beta^2(3 - \lambda^2)
	 -  4\beta\lambda +3\lambda^2-1 
          } ,
\label{eqn:beta}
\ee
and the density is given as a simple integral over the fluid velocity,
\be
{\tilde \rho}={\tilde
\rho}_s\exp\left\{\int_\lambda^{\lambda_s}\frac{8\beta\left(
\lambda' -\beta  \right)d\lambda'}{  \beta^2(3 - \lambda'^2)
	 -  4\beta\lambda' +3\lambda'^2-1} 
\right\}.
\label{eqn:n}
\ee
This self-similar scaling imposes a constant shock speed with velocity
$\lambda_s$.  The subscript $(...)_s$ on a fluid variable denotes its value
immediately behind the shock front, e.g.  ${\tilde
\rho}_s\equiv\lim_{\lambda\rightarrow\lambda_s^-} {\tilde \rho}(\lambda)$.

In order to solve the system of equations (\ref{eqn:beta}) and
(\ref{eqn:n}), we need to specify the postshock boundary conditions.
Equations~(\ref{eqn:cont}) and~(\ref{eqn:momt}) provide the postshock
velocity in the star rest frame and the postshock density in the fluid
frame:
\begin{eqnarray}
\beta_s&=&{\frac{2[  {{\lambda_s}^2}( 1 - {\tilde \rho}_V )  - \alpha-
{\tilde \rho}_V ]
          - {\sqrt{-3{{( 1 + \alpha ) }^2}{{\lambda_s}^2} + 4{{[ \alpha -
{{\lambda_s}^2}( -1 + {\tilde \rho}_V ) + {\tilde \rho}_V ] }^2}}}}
{\lambda_s ( 3 - \alpha - 4{\tilde \rho}_V 
) }},
\label{eqn:betas}
\\
{\tilde \rho}_s&=&{\frac{3( 1 + \alpha ) 
      ( 1 - \beta_s^2) \lambda_s}{4
      ( \lambda_s -\beta_s) 
      ( 1 - \beta_s\lambda_s ) }}.
\end{eqnarray}
Requiring the postshock velocity to be real in the shock frame
yields the inequality 
\be
\frac{3+\alpha(3\alpha-2)+8{\tilde \rho}_V({\tilde \rho}_V+\alpha-1)+
\sqrt{3}(1+\alpha)\sqrt{(4
{\tilde \rho}_V+\alpha-3)(4{\tilde \rho}_V+3\alpha-1)}}{8(1-
{\tilde \rho}_V)^2} < \lambda_s^2 < 1
\ee
which simplifies to $3/4<\lambda_s^2<1$ if $\alpha={\tilde \rho}_V=0$ in
the idealized case of a low-pressure neutron star and a negligible burning
potential.  We conclude that the burning front must propagate
supersonically in the star frame.  In the shock frame, the post-shock
velocity of the fluid is smaller than its pre-shock value, therefore
this solution corresponds to a detonation front.
%In the fluid frame the shock also moves supersonically, 
%approaching the speed of sound as the star frame shock
%speed approaches its lower boundary.  
By examining equation (\ref{eqn:beta}) we find that all solutions must pass
through the point $\beta=0$, $\lambda=1/\sqrt{3}$.  The only physical
solution in the range $0\le \lambda \le 1/\sqrt{3}$ is $\beta=0$, so any
solution is matched continuously at the boundary with discontinuous
derivatives.  Normally one specifies a boundary condition away from the
shock front, but in our case this boundary condition is automatically
satisfied.  The shock speed $\lambda_s$ is therefore a free parameter which
can lie anywhere in the allowed range.  The actual value of $\lambda_s$ is
either set by the trigger mechanism (i.e., the initial conditions) or by
the viscous forces acting on the domain wall boundary.  In the generic
case where viscous forces slow down the shock to its minimum velocity,
the unique solution is
\be
\lambda_s^2=\frac{3}{4}-\frac{\alpha}{2}-\frac{{\tilde \rho}_V}{2} ,
\ee
to leading order in $\alpha$ and ${\tilde \rho}_V$.  The postshock density
is $n_s\simeq 3-4\alpha-7{\tilde \rho}_V$ and the exact postshock velocity
\be
\beta_s^2=1-\frac{2(1+\alpha)}{3-\alpha-4{\tilde \rho}_V} ,
\ee
giving the constraint $3\alpha+4{\tilde \rho}_V<1$.  Even though the
velocity derivative $\beta'$ formally diverges, the solution is well
defined.

\begin{figure}
\plotone{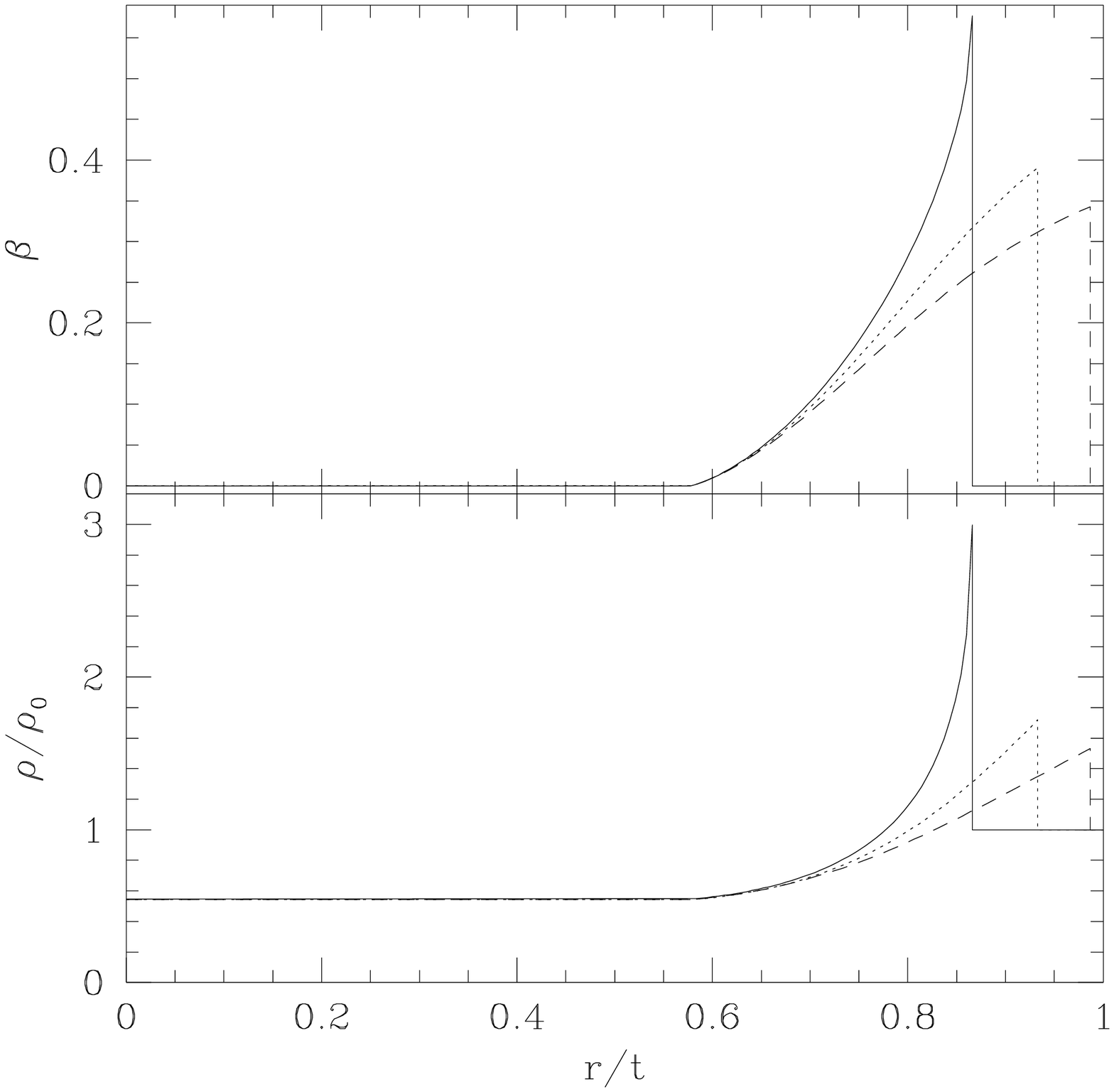}
\caption{Self-Similar explosion solution for $\alpha=0$
and ${\tilde \rho}_V=0$.  The different curves correspond to different
detonation front speeds, $\lambda_s$.  The solid line is the generic
solution in the presence of weak viscosity.}
\label{fig:blast}
\end{figure}

By virtue of equation~(\ref{eqn:momt}),
all shock solutions automatically satisfy total energy conservation
\be
\frac{T^{00}}{\rho_0t^3}=\frac{4}{3} \pi \lambda_s^3 = \frac{4}{3} \pi
{\tilde \rho}(\frac{1}{\sqrt{3}}) 3^{-3/2} + \frac{4}{3} \pi \lambda_s^3
{\tilde \rho}_V + 4\pi
\int_{1/\sqrt{3}}^{\lambda_s}
\left(\frac{1+\beta^2/3}{1-\beta^2}\right){\tilde \rho}\lambda^2 d\lambda ,
\ee
which provides a consistency check for the final numerical solution.
Figure~\ref{fig:blast} shows the numerical solutions for the velocity and
density profiles behind the shock for three different choices of
$\lambda_s$, assuming $\alpha=0$ and ${\tilde \rho}_V=0$.

\section{Trigger Mechanism}
\label{sec:ew}

We have found that if triggered, a self-sustained burning front would
propagate at a constant speed and spontaneously convert baryons at nuclear
density into radiation.  However, we still need to demonstrate how this
process might be spontaneously triggered, and under which conditions the
burning front would release sufficient energy to sustain itself.

The rate of neutron star mergers in the Hubble volume is similar to the
observed rate of GRBs (Narayan et al. 1991; Phinney 1991).  It is
therefore natural to examine the question whether an electroweak burning
front would be triggered by such mergers.  Typically, the combined mass of
the two neutron stars would exceed the maximum mass for a stable star at
nuclear densities, and the merger product would be gravitationally unstable
and tend to collapse to a black hole. An electroweak burning front might
be produced if there is an inner region where the density is
sufficiently high to restore electroweak symmetry before being engulfed by
a black hole horizon.  In this region, baryon-number will be strongly
violated and most baryons will decay into anti-leptons.
 Similar conditions might occur in binary systems where a
neutron star exceeds the maximum stable mass due to accretion from its
companion. However, the formation of black holes by much more massive
systems, such as the direct collapse of massive stars, are characterized
by much lower densities (at the Schwarzschild radius, $\rho\propto 1/M^2$)
and are not expected to trigger the electroweak transition before their
horizon forms.

The details of the trigger process depend on the equation of state of
neutron star matter above nuclear density, which is not yet fully
understood.  Some nuclear equations of state predict that $p>\rho/3$ for
densities about ten times nuclear (Prakash et al. 1997).  The sound speed
$c_s^2=\partial p/\partial \rho$ then obtains a maximum value above that of
a pure radiation fluid.  At yet higher densities, the sound speed
eventually declines back to the $1/3$ value of a locally interacting
radiation field due to asymptotic freedom of QCD.  If we now consider a
neutron star which collapses slowly through phases of hydrostatic
equilibrium, then the central equation of state will stiffen until it
reaches its maximum sound speed, and then soften again.  The region which
enters the softened phase will collapse faster than its surroundings,
potentially leading to a runaway density.  The compression of this region
is caused by the weight of the column of matter above it.  The electroweak
symmetry would be restored in this region at a density $\sim 10^{27}$
g/cm$^3$, some twelve orders of magnitude above the initial density.  This
region will not be surrounded by a black hole horizon if its radius is $\la
1~{\rm cm}$, or its mass is $\la 4\times 10^{-6}M_\odot$.  A lower bound on
the radius of the trigger region arises from the requirement that a
self-sustained burning front needs to release more energy interior to it
than the total energy stored in the domain wall.  At nuclear density the
energy released is $\sim (200 {\rm MeV})^4 R^3$ while the wall energy is
$\sim(250 {\rm GeV})^3 R^2$, implying a radius $\ga 10^{-3}$ cm. Higher
background densities will only weaken this constraint.

We have two apparently contradictory requirements.  The fluid must collapse
to electroweak densities, and then expand again to drive the explosion.
In the dense phase, the relative pressure gain from the burning of each
nucleon is small.  One would expect that the gravitational collapse and the
subsequent phase transitions will raise the entropy of the fluid in the
central region and cause entropy inversion, therefore resulting in
convective flows.  A fluid element taken through the electroweak phase
transition which would subsequently rise due to buoyancy forces, could
trigger the explosion as it cools adiabatically and is trapped in
the false vacuum.

\section{Kinematics of Baryon Number Violation on Propagating Walls}
\label{sec:energ}
In this section we discuss the energetics and timescales of the baryon
number violating processes on bubble walls moving through cold neutron
matter.  We consider the possibility that neutrons are converted into
antineutrinos as the bubble wall passes. Our analysis will be extremely
schematic, at best a pointer to the key issues involved. At the end of this
section we shall comment on the additional work which is needed in order to
settle the microphysics of electroweak bubble walls in neutron matter.

We shall consider particle physics models which include more than one Higgs
doublet field. These models are amongst the most plausible extensions of
the standard model and occur as sub-models of low energy supersymmetry and
attempts to incorporate generational symmetries in the standard model
(Gunion et al. 1990).  We are particularly interested in these models
because they possess stable domain wall solutions within which the Higgs
fields may be very small. The domain wall solutions arise as an inevitable
consequence of discrete symmetries imposed to prevent phenomenologically
unacceptable flavor changing neutral currents.  For example, if the field
$\phi_U$ gives mass to up quark and $\phi_D$ to the down quark, one imposes
a symmetry: $\phi_U \rightarrow -\phi_U$, $U_R\rightarrow-U_R$, with $U_R$
being the right-handed up quark. This prevents tree level processes in
which neutral Higgs intermediate states allow up type quarks to change into
down type quarks (Glashow \& Weinberg 1977). Of course these symmetries
mean that at least at the classical level the theories possess degenerate
vacua, and the domain walls are stable classical field configurations
interpolating between them.

The domain walls are interesting to us because in their interior the
gauge field masses may be very small and the barrier to baryon number
violating processes is therefore small.  Furthermore, as pointed out
by Preskill et al. 1991, in these particular theories the domain walls
pose no problems for cosmology, since they vacuum degeneracy is
actually broken by very small instanton induced effects, the vacuum
energy being different by $\sim {1\over 3} f_{\pi}^2 m_{\pi}^2 \sim
(80$ MeV$)^4$.  They may in fact be useful for producing a
cosmological baryon asymmetry, as we discuss below.

The electroweak baryon number violating processes violate baryon and lepton
number, but conserve their difference.  A unit change in the gauge/Higgs
winding number destroys 3 baryons and 3 leptons. Since quarks, but not
leptons, are allowed to change flavors through Cabibbo-Kobayashi-Maskawa
mixing, one can have the processes $3n
\rightarrow \overline{\nu_e}+ \overline{\nu_\mu}+
\overline{\nu_\tau}$, or $2n \rightarrow \overline{n}+
\overline{\nu_e}+ \overline{\nu_\mu}+
\overline{\nu_\tau}$.
A key prediction of our mechanism is that the high energy neutrinos 
measured from these processes would exhibit an equal 
excess of antineutrinos from each generation.

The basic equation governing the rate of baryon number violation
in the standard model reads,
\be
{dn_B\over dt} = {{\cal N} \over 8\pi^2} (g_2^2 {\bf E\cdot B} -g_1^2 {\bf
e \cdot b}) ,
\label{eq:anom}
\ee
where $n_B$ is the baryon number density, ${\cal N} =3$ the number of
generations, $g_2$ and $g_1$ the $SU(2)$ and $U(1)_Y$ gauge coupling
constants and ${\bf E,B,e,b}$ the corresponding electric and magnetic
fields. The linear combination occurring on the right hand side of
(\ref{eq:anom}) vanishes for electromagnetic fields: baryon number
violation requires the $W$ and $Z$ fields. If the wall moves at velocity
$v_{\rm w}$, then the number of neutrons hitting the wall per unit time per
unit area is $\gamma_{\rm w} n_N v_{\rm w}$, where $n_N$ is the ambient
neutron density. Requiring that a sizeable fraction of these neutrons are
converted then leads to an estimate of the required electric and magnetic
fields.

Assuming equipartition between electric and magnetic fields, and a
semi-relativistic wall velocity, we find the needed fields are:
\be
B_{required} \sim E_{required} \sim \sqrt{\gamma_{\rm w} n_N v_{\rm w}
m_H/\alpha_2}
\sim(1.5{\rm GeV})^2 ,
\label{eq:fields}
\ee
with $\alpha_2 = g_2^2/4 \pi \sim {1\over 30}$ the weak fine structure
constant. Here we have integrated the anomaly across the wall width, taken
to be the inverse Higgs mass $m_H \sim 100$GeV, and have ignored $g_1$
since it is much smaller than $g_2$.  The resulting fields are indeed
large, but they are very much smaller than the electroweak scale. We have
ignored the Cabibbo-Kobayashi-Maskawa mixing angle suppression.  The B
violating processes would presumably create heavy quark states which would
be trapped upon the wall since they are light on it, which would annihilate
with ups and downs via CKM mixing.

Could the required fields be accreted on the wall?  It is clear that the
physics of the domain walls is quite complex and there are many possible
contributing effects.  Fermions are attracted to the wall, since their
masses are lower on it, and form bound states which may support currents
flowing without resistance on the wall.  Since the neutrons carry weak
isospin, their presence automatically causes a weak isospin electric fields
of order $g n_N^{2\over 3}$ on the wall if the gauge fields are massless
there. But the most likely mechanism for generating the required $\sim$ 1
GeV$^2$ fields is the Meissner effect -- $W$ and $Z$ gauge fields
would be swept along by the back edge just because they are light on the
wall and massive off it. It is also possible that a dynamo mechanism would
operate on the wall, building up the magnetic fields to the saturation
value $B_{\rm crit}\sim (100 {\rm GeV})^2$.

How much energy do the fields in equation (\ref{eq:fields}) cost? And
how does this compare with the rate of energy input in the form of
neutrons onto the wall? Equating the energies per unit area
$B_{required}^2/m_H \sim n_N L_N m_N$ where $L_N$ is the path swept up
in the cold neutrons, we find that the required field energy could be
accreted in a distance as short as $\sim 3$ neutron spacings at
nuclear density, or $L_N \sim (\alpha_2 m_N)^{-1}$.  So there is
certainly no shortage of energy input to maintain the fields.

We require aligned electric and magnetic fields in order to violate
baryon number: we may think of the process occurring in two stages.
First magnetic fields are established on the wall via some kind of
dynamo mechanism coupled to the Meissner effect.  In a constant
$SU(2)$ magnetic field, the lowest energy levels for the quark
consists of left handed particles with positive isospin (up quarks)
moving against the field and left handed particles with negative
isospin (down quarks) moving along the field.  Right handed particles do
not couple to the field.  A beam of neutrons streaming into such a field
would undergo isospin separation, leading to an electric field aligned
with the magnetic field.  The resulting change in the gauge field
winding causes neutrons to disappear.

Some elementary steps towards establishing the viability of this mechanism
would be: (i) finding the energy barrier required for baryon number
violation in the presence of a domain wall, (ii) studying the fate of
currents on the wall, and whether they can support magnetic fields of the
required magnitude, and finally (iii) studying the possibility of a dynamo
mechanism that would establish such fields out of a seed magnetic field,
that is likely to be present in the neutron star. We shall return to these
questions in future work.

Assuming the electroweak baryon number violation processes are sustained in
the bubble wall, the latter will be pushed forward through the neutrons by
the excess pressure of the hot medium behind it. Presumably it will reach a
terminal speed determined by collisions with the neutron matter.  The
self-sustained detonation front described in \S 2 could convert all the
nuclear matter mass $M_{\rm tot}$ into neutrinos and photons, releasing
$\sim 5\times 10^{54}~{\rm ergs}\times (M_{\rm tot}/3M_\odot)$ over a time
$\la (30~{\rm km}/c)\sim 10^{-4}~{\rm sec}$.

When the electroweak domain wall emerges from the surface of the
neutron star, the domain wall will still expand due to the thermal
pressure behind it out to $R_{\rm max}/R_{\star}\sim {\tilde
\rho}_V^{-1/3}$ where $R_\star$ is the radius of the neutron star. The
lack of additional fuel will prevent it from expanding further; beyond this
radius, the radiation fireball will diffuse out and separate from the
domain wall.  This energy release will be the primary source of energy for
the GRB.  As discussed in the next section, typically ${\tilde \rho}_V \sim
1/100$, and the limiting radius would be $R_{\rm max}\sim 10^{7}$ cm.  If
${\tilde \rho}_V=0$ the wall will not expand forever, but rather at $R\sim
10^{10}$ cm it will contract again due to wall tension.  Eventually, the
wall would lose the radiation pressure support behind it and collapse
again.  The Higgs energy of the wall and the false vacuum behind it will be
radiated as high energy ($\sim 100~{\rm GeV}$) neutrinos, photons, and
particle-antiparticle pairs.  In contrast to supernovae, where neutrinos
carry 99\% of the energy (since photons are trapped by the opaque stellar
envelope), there should be comparable energies in neutrinos and photons in
the resulting fireball here.  In addition to the $\sim 10^{54}$ ergs
radiated in neutrinos of energy $\sim$ 100 MeV, a second sub-millisecond
burst carrying $\sim 10^{52}$ ergs of $\sim 100$ GeV neutrinos is a generic
prediction of our model.

\section{Implications for Electroweak Theory and Cosmological Baryogenesis} 
\label{sec:cosm}

\def\gga{\mathrel{\hbox{\rlap{\hbox{\lower4pt\hbox{$\sim$}}}\hbox{$\gg$}}}}

Electroweak domain walls can be cosmologically disastrous.  When formed in
a first order phase transition, they will come to dominate the matter
density of the universe (Zel'dovich et al. 1975).  If we assume about one
domain wall per horizon volume, electroweak domain walls would dominate the
matter density of the Universe if they persist until its temperature is
$\sim 1$ keV.  Thus, there is a cosmological lower bound on the potential
difference $\Delta V \equiv V^--V^+ \gg (1~{\rm keV})^4$.  Preskill et al.
(1991) argued that discrete symmetries could be broken anomalously through
QCD effects, yielding an energy difference $\Delta V \sim (80$ MeV$)^4$.
The vacuum energy admits an equation of state $p=-\rho$.  The energy
density of a neutron star is $\sim (200$ MeV$)^4$.  Requiring the postshock
pressure to be positive, we obtain an upper bound to this energy asymmetry
$\Delta V\la (150$ MeV$)^4$, in order for the burning front to be
self-sustained. This bound is well above the value predicted by Preskill et
al.  (1991).  We also see that such electroweak explosions can only occur
in matter which has densities close to nuclear, i.e. only in neutron stars.

A related enigma in cosmology is baryogenesis.  Baryon number is
strongly violated above the electroweak phase transition, and in some 
theories (where baryon number minus lepton number is conserved)
baryon asymmetries generated before this transition would be erased.
Baryon number is conserved after the electroweak symmetry breaking,
so the natural time to create baryon number would be during this phase
transition.  The Sakharov criteria require that baryon number B,
charge conjugation symmetry C, and parity CP be violated, and that the
system be out of equilibrium.  In the minimal standard model of
particle physics, two of these criteria are problematic.  The current
measurement of the top mass $m_{\rm top} \sim 180$ GeV appear to preclude a
first order phase transition, which would be the natural way to bring
the Universe out of equilibrium.  Also, the magnitude  of CP violation in
the minimal standard model cannot produce the observed
baryon-to-photon ratio $\eta \sim 10^{-10}$.  Moving topological
defects such as domain walls restore the out-of-equilibrium
requirement (Prokopec et al. 1996), and the extra freedom in selecting
the second Higgs field also allows a match for the required CP
violation rates.  The long persistence of the anomalous domain walls
will also amplify the net baryon production to become more consistent
with the small observed CP violation rates.  

Another consequence of the long persistence of these walls is a
coherence length of the baryon production regions which is
intermediate to the horizon size at electroweak and the horizon size
when the domain walls disappear.  This length scale of $\sim
10^{15}$--$10^{19}$ comoving cm, is significantly longer than the
diffusion length at nucleosynthesis (Jedamzik and Fuller 1995).  The
baryon-to-photon ratio would generically fluctuate in different
regions, resulting in inhomogeneous nucleosynthesis and affecting the
standard predictions for the abundance of light elements.

\section{Conclusions}
\label{sec:conc}

We have argued that the gravitationally unstable core of a neutron
star merger might trigger an electroweak burning front which would
propagate outwards and convert all the nuclear material into
radiation.  The associated detonation front would be self-sustained
and propagate at a constant semi-relativistic speed $\sim
\sqrt{3/4}c)$.  The baryon-number violating reaction is induced by a
domain wall, which is naturally predicted in double Higgs models of
electroweak interactions.  We have shown that the electroweak burning
front can only be self-sustained if it propagates into material with
nuclear densities.  Although our mechanism provides an efficient way
for alleviating the baryonic contamination problem, some surrounding
stellar debris might remain unburnt and limit the Lorentz Factor of
the expanding fireball. Minimal contamination is required to make the
emission spectrum from the fireball non-thermal (Goodman 1986;
Paczy\'nski 1986).

The electroweak explosion process releases $\ga 10^{54}$ ergs over
$\la~{\rm msec}$, and produces a fireball which subsequently impacts on the
ambient (interstellar) medium.  This impact generates an external shock
which could produce via synchrotron emission and inverse-Compton scattering
of Fermi-accelerated electrons, the observed spectrum of GRBs (M\'esz\'aros
\& Rees 1993a,b; M\'esz\'aros, Laguna, \& Rees 1993).  
In this model, the observed variability of GRBs might be either due to
inhomogeneities in the ambient gas density, or due to instabilities at the
shock front (e.g.  Waxman \& Piran 1994).  The release of $\ga 10^{54}$
ergs remedies the energy problem which was previously identified for
external shock models (Sari \& Piran 1997a,b), and is in fact required if
GRBs follow the star formation history of the Universe with the dimmest
bursts originating at redshifts as high as $z\sim 6$ (Wijers et al.  1997).
The integrated light from afterglow observations can set a lower bound on
the total energy released, and in principle challenge models which provide
too little energy.  The release of $\sim 10^{54}~{\rm ergs}$ in more
conventional astrophysical settings (see e.g., the hypernovae model of
Paczy\'nski 1997) is likely to be contaminated by baryons at a level which
depends on many random parameters, such as the orientation of line-of-sight
through the source.  Under such circumstances, one would expect a
continuous distribution of events with varying levels of baryonic
contamination, from the level of $\la 10^{-2}M_\odot$ required for GRBs and
up to the level of $\sim 10M_\odot$ found in supernovae. Such models would
therefore predict the existence of softer events, such as X-ray bursts
(which were not observed so far), in addition to GRBs. Our model naturally
accounts for the {\it gap} between the baryonic contamination levels found
in GRBs and supernovae.

The collapse of the bubble wall could also release a fraction
$\tilde\rho_V $ of the neutron star rest mass, i.e. $\sim 10^{52}$
ergs, when some of the Higgs potential energy is converted into
non-thermal neutrinos with very high energies $\sim 100$ GeV.  The
number of $\sim 100~{\rm GeV}$ neutrinos per unit area from a source
at a distance $\sim 3~{\rm Gpc}$ would be $\sim 10^{6}~{\rm km}^{-2}$.
Full timing and directional coincidence with GRB events might allow
statistical detection by large area Cherenkov arrays, such as AMANDA.
With a typical conversion efficiency of $10^{-8}$ into upward moving
muons (Gaisser et al.  1995), a year long cross-correlation with $\sim
300$ bursts could yield $\sim 3$ events with a km$^2$ array.  This
neutrino burst supplements the predicted neutrino flux from standard
fireball models (Waxman \& Bahcall 1997, Paczy\'nski and Xu 1994).  A
larger amount of energy $\sim 10^{54}$ is released in $\sim 200~{\rm
MeV}$ neutrinos, remnant of the burnt phase behind the detonation
front; however, these lower energy neutrinos would be more difficult
to detect.

Finally, we note that the extended Higgs models are amongst the simplest
extensions of the standard model, and the discrete symmetries needed to
produce the domain walls are required by particle physics phenomenology.
These models may well provide a new solution to the problem of electroweak
baryogenesis, with associated inhomogeneous cosmological nucleosynthesis.
The quantitative constraints which are placed on the double Higgs theory by
our GRB model could be tested with future accelerator experiments.
Conversely, if our proposed link between electroweak theory and GRB is
established, GRBs might be used to place constraints on parameters of the
Higgs sector which go beyond the capabilities of laboratory accelerators.

\acknowledgements

We thank Shmulik Balberg and Eli Waxman for useful discussions.  This
work was supported in part by the NASA ATP grant NAG5-3085 and the
Harvard Milton fund (for AL and UP).

\end{document}